Strain tunability of perpendicular magnetic anisotropy in van der Waals ferromagnets $VI_3$

Xi Zhang[1,2+], Le Wang[2,3+], Huimin Su[2,3], Xiuquan Xia[2], Cai Liu[2,3], Junhao Lin[4], Mingyuan Huang[4], Yingchun Cheng[5*], Jia-Wei Mei[4,6*], Jun-Feng Dai[2,3,7*]

1. Frontiers Science Center for Flexible Electronics (FSCFE), Xi'an Institute of Flexible Electronics (IFE) and Xi'an Institute of Biomedical Materials & Engineering (IBME), Northwestern Polytechnical University, Xi'an 710072, China
2. Shenzhen Institute for Quantum Science and Engineering, Southern University of Science and Technology, Shenzhen, 518055, China
3. International Quantum Academy, Shenzhen, 518048, China
4. Department of Physics, Southern University of Science and Technology, Shenzhen, 518055, China
5. Key Laboratory of Flexible Electronics & Institute of Advanced Materials, Jiangsu National Synergetic Innovation Center for Advanced Materials, Nanjing Tech University, Nanjing 211816, China
6. Shenzhen Key Laboratory of Advanced Quantum Functional Materials and Devices, Southern University of Science and Technology, Shenzhen 518055, China
7. Shenzhen Key Laboratory of Quantum Science and Engineering, Shenzhen 518055, China

+ The authors contribute to this work equally
* Corresponding authors:
daijf@sustech.edu.cn; meijw@sustech.edu.cn; iamyccheng@njtech.edu.cn

**Abstract**

Layered ferromagnets with high coercivity have special applications in nanoscale memory elements in electronic circuits, such as data storage. Therefore, searching for new hard ferromagnets and effectively tuning or enhancing the coercivity are the hottest topics in layered magnets today. Here, we report a strain tunability of perpendicular magnetic anisotropy in van der Waals (vdW) ferromagnets $VI_3$ using magnetic circular dichroism measurements. For an unstrained flake, the M-H curve shows a rectangular-shaped hysteresis loop with perpendicular magnetic anisotropy and a large coercivity (up to 1.775 T at 10 K). Furthermore, the coercivity can be enhanced to a maximum of 2.6 T at 10 K under a 2.9% in-plane tensile strain. Our DFT calculations show that the magnetic anisotropy energy (MAE) can be dramatically increased after applying an in-plain tensile strain, which contributes to the enhancement of coercivity in the $VI_3$ flake. Meanwhile, the strain tunability on the coercivity of $CrI_3$, with a similar crystal structure, is limited. The main reason is the strong spin-orbital coupling in $V^{3+}$ in $VI_6$ octahedra in comparison with that in $Cr^{3+}$. The strain tunability of coercivity in $VI_3$ flakes highlights its potential for integration into vdW heterostructures, paving the way toward nanoscale spintronic devices and applications in the future.

The study on layered magnets coupled by vdW force can be traced back to 50 years ago[1-3], however, a renewed interest has been driven recently due to the discovery of magnetism in vdW ferromagnets down to the two-dimensional (2D) limit[4-8]. This opens a new era of developing 2D spintronic devices for electronic applications based on magnetic 2D materials. For a ferromagnet, coercivity refers to the opposing magnetic intensity that is applied to a ferromagnetic material to remove the residual magnetism. On the other hand, it reflects the ability of ferromagnetic materials in retaining remanence. Hard ferromagnetic materials, generally having large coercivity, indicate great potential for applications in the manufacture of magnetic storage units. However, for most well-known layered ferromagnets, coercivity is generally less than one tesla (T) [$CrI_3$ (~0.2 T) and $Fe_3GeTe_2$ (~0.6 T)[9]]. This seriously weakens their ability to resist external magnetic interference. Therefore, this stimulates efforts to discover new 2D magnets with high coercive force or elevate the coercivity in the uncovered 2D magnets. In comparison with 3D bulk ferromagnets, the ferromagnetic properties of vdW magnets are sensitive to external stimuli, such as hydrostatic pressure and strain[10-12]. It will directly act on the inter- or intra-atomic distance and the bond angle between transition atom and ligand in the crystal, leading to the tunability of the coercivity[13, 14]. Although the percentage of coercivity dramatically increases in most 2D magnets by an external force, the absolute value is still less than 1 T. Generally speaking, hard ferromagnetic materials have very strong magneto-crystalline anisotropy, which comes from a joint effect of crystal field, spin-orbit coupling (SOC) and super-exchange interaction between spins. Therefore, one strategy is to apply external forces on the vdW magnets with strong MAE, tuning the coercivity by the strong spin-orbital interaction. Here, we conduct a comparative experiment on strain tunability of coercivity in vdW ferromagnets $VI_3$ and $CrI_3$, with and without the orbital moment[15, 16], respectively. We found that ~2.9% tensile strain in the layered ferromagnet $VI_3$ can induce a coercivity of 2.6 T at 10 K, with a growth of 46% in comparison with the unstrained case. Our DFT calculations indicate the MAE dramatically increases as the tensile strain increases from 0 to 2%, which contributes to the enhanced coercivity under tensile strain. In contrast, $CrI_3$, with a similar crystal structure as $VI_3$, shows a limited modulation of the coercivity by the tensile strain due to the quenching orbital moment. Our experimental results justify that $VI_3$ is a potential candidate for developing magnetic data storage and functionalized spintronic devices.

**Sample characteristics**

$VI_3$ has a layered honeycomb structure coupled with vdW force (Fig. 1a)[17]. Its single crystal can be grown by a chemical vapor transport method (see Methods for details). The representative sample size is around 2 × 2 $mm^2$ as shown in the inset of Fig. S1. The X-ray diffraction (XRD) results in Fig. S1(a) indicate the high quality of the sample. The magnetic susceptibility ($\chi$) measurement in Fig. S1(b) and (c) reveals a structural transition at $T_s$ of 78.5 K and a ferromagnetic phase transition at $T_c$ of 49.4 K with an easy axis along the crystallographic c-axis. Under the maximum applied field of 7 T and at 1.8 K (Fig. 1b), the magnetic moments are saturated as 1.156 $\mu_B/V^{3+}$ for H // ab and 1.367 $\mu_B/V^{3+}$ for H // c, which are both smaller than the expected saturated moment of a $V^{3+}$ ion, gS = 2 μB/f.u. Moreover, the out-of-plane curve shows a coercivity field of 1.2 T, much bigger than that of the in-plane (0.1 T), which again suggests the easy axis to lie along the *c* axis.

**Experimental results**

We first characterize the basic magnetic property of exfoliated $VI_3$ flakes on a $SiO_2$/Si substrate using temperature-dependent magnetic circular dichroism (MCD) measurement. Fig.1c shows a representative field-dependent magnetization curve (M-H curve) in a 27 nm $VI_3$ flake (Fig. S2) at several fixed temperatures, with an external magnetic field along the *c*-axis of the sample. At low temperature, the hysteresis loop shows a rectangle shape, indicative of an out-of-plane ferromagnetic order, which is consistent with the field-dependent magnetization measurement in the $VI_3$ single crystal. As temperature increases, the hysteresis loop gradually shrinks and eventually closes at 56 K, and then changes to a linear relationship between magnetization and applied field above 60 K, indicating a transition from ferromagnetic to paramagnetic states. This temperature-dependent magnetic property is consistent with the magnetic susceptibility measurement of the single crystal. However, the Curie temperature ($T_c$) is higher than 49.5 K obtained from magnetic susceptibility measurement, which is related to the thickness-dependent magnetism as reported in Ref. [18]. While, the averaged coercivity extracted from MCD measurement is around 1.48 T at 10 K, which is one order of magnitude larger than the isostructural compound $CrI_3$ (Fig. 1d). It is probably due to the additional contribution of the orbital moment of $V^{3+}$ ion in $VI_3$ as discussed in the Ref. [19], whereas, the orbital angular momentum is quenched in $CrI_3$ [16]. It should be noted that there is a negative shift in the M-H curve (Fig. 1d), despite a rectangle hysteresis loop. This phenomenon can be found in most $VI_3$ flakes. This can be easily understood as the presence of antiferromagnetic orders between layers, which pin the direction of ferromagnetic order due to the interfacial exchange coupling between ferromagnetic and adjacent antiferromagnetic layers. This phenomenon has also been reported in other 2D magnets such as $Fe_3GeTe_2$[20]. To better evaluate the effect of strain on coercivity, we selected the $VI_3$ flakes showing a ferromagnetic ground state to conduct the following experiments.

Then, we dynamically apply an in-plane tensile strain in a thin $VI_3$ flake and detect the temperature-dependent ferromagnetic properties using MCD spectroscopy simultaneously. To maximize the tensile strain, we select a relatively thin $VI_3$ (~20 nm) as a candidate. As shown in Fig. 2a, we employ polymer polydimethylsiloxane (PDMS)/Si substrate as a stretcher and transfer the thin $VI_3$ flake on it by dry transfer method, then gradually stretched the PDMS film to exert a tensile strain on the sample. Then the strained sample was fixed to a cold finger of the low-temperature cryostat to conduct MCD measurements (Fig. S3). Fig. 2b and Fig. S4 show the optical image of the $VI_3$ flake before and after stretching. By measuring the length difference between them, we can determine a maximum of ~2.9% tensile strain applied on the sample while preventing sample breakage. The experimental method in detail can be found in the Method.

Fig. 2c shows the normalized polar MCD curves at a fixed temperature of 10 K on the same point of a $VI_3$ flake when the in-plane tensile strain increases from 0 to 2.9%. Without strain, the hysteresis loop shows a symmetrical rectangle shape around the zero magnetic field with a coercivity of 1.78 T at 10 K, indicating a very weak interlayer antiferromagnetic coupling in the thin sample. The MCD mapping at 10 K near the transition field is shown in Fig. 2b, where the direction of magnetization of the entire sample at the unstrained state changes when the applied magnetic field increases from 1.75 T to 1.80 T. The uniformity of coercivity reflects negligible microscopic strain within the flake. In addition, the Curie temperature is extracted to be around 64±2 K (Fig. 3a). As the strain increases, we could find that a ferromagnetic ground state is still preserved. Surprisingly, the coercive force gradually increases as the tensile strain increases, with a maximum of 2.60 T at 10 K under a 2.9 % strain (Fig. 2c). Under strain, the range of transition field, which corresponds to a magnetic field to change the direction of magnetization, is within 0.05 T at 1.7% strain and 0.1 T at 2.9% strain (Fig. 2b), respectively. It also reflects that the applied tensile strain is nearly uniform within the entire sample. As the strain further increases, the $VI_3$ flake tends to crack, accompanied by a decrease in coercivity because of strain release (Fig. S5). Moreover, the strain-dependent MCD results in another thin $VI_3$ flake, as shown in Fig. S6 and S7, also show an enhancement of the coercivity as the strain increases by 2% strain. However, the temperature-dependent MCD measurement shows that the Curie temperature does not change as the strain increases (Fig. 3a and Fig. S4). The strain-related enhancement of coercivity also fades with the increase in temperature (Fig. 3b). This may be related to another ferromagnetic phase below $T_c$, where two magnetically ordered V sites contribute to the magnetic order below 36 K, accompanied by a structure transition[19, 21]. This ferromagnetic state is very sensitive to external stimuli, such as pressure[22]. So far, we have achieved an effective enhancement of coercivity in the thin $VI_3$ flake up to 2.6 T at 10 K through a 2.9% in-plane tensile strain.

For the $V^{3+}$ in $VI_3$, two 3d electrons occupy two of the three degenerate $t_{2g}$ electronic states in its $VI_6$ octahedra with the total spin S=1. Due to partially filled $t_{2g}$ orbitals, it leads to the presence of orbital degrees of freedom in $VI_3$[23, 24], which is antiparallel to its spin moment along the *c*-axis. It also indicates that the SOC will be much stronger in VI$_3$ in comparison with other vdW ferromagnets with quenching orbital moments such as CrI$_3$. It is the orbital motion of electrons in V$^{3+}$ ion-based ferromagnets that strongly couples with a crystal electric field, leading to its contribution to strong magneto-crystalline anisotropy. The theoretical study in literature also justifies that MAE in VI$_3$ is an order of magnitude larger than that in CrI$_3$ [16]. These are the primary origins of hard ferromagnetism in VI$_3$, namely, high coercivity. However, under tensile strain, how the subtle distortion of VI$_6$ octahedra induces the enhancement of coercivity is still uncertain. Therefore, we employ DFT calculations to study the in-plane tensile strain effect on MAE in monolayer VI$_3$. Fig. 4a illustrates the direction of magnetic moment ($\mu_s$) using θ and α, where the former is defined as the angle between $\mu_s$ and *c*-axis and the latter is that between the projection of $\mu_s$ in *ab*-plane and *a*-axis. As shown in Fig. 4b, the in-plane MAE is nearly isotropic in both the 1% strain and unstrained cases, which means that the uniaxial tensile strain does not change the isotropy. Without strain, the magnetic easy axis is along the *c*-direction, which is consistent with calculated results in literatures[16, 25]. After applying a 1% uniaxial tensile strain, the easy axis remains along the *c*-axis (Fig. 4c). More importantly, the MAE increases as the uniaxial tensile strain increases from 0 to 2%. The increase of MAE well explains the enhanced coercivity under tensile strain. However, the experimentally observed T$_c$ of VI$_3$ is insensitive to the tensile strain (Fig. 3), which contradicts the previous theoretical result[24]. Therefore, more theoretical investigation deserves to explore the dependence of T$_c$ on the strain, but it is beyond scope of current work.

In the bulk form, CrI$_3$ also shows a ferromagnetic behavior with the easy axis along the *c*-axis (Fig. S8). In contrast to VI$_3$, the Cr$^{3+}$ ion in CrI$_3$[26] has three 3d electrons, occupying the half electronic states in t$_{2g}$ with zero orbital angular momentum. It means that SOC is weak, indicating a relatively small MAE. To check the tunability of coercivity in CrI$_3$ under a tensile strain, we also perform the strain-related MCD measurements in two fixed points of a multilayer sample at different temperatures (Fig. 5). For point 1, without strain, the M-H curve in CrI$_3$ shows a standard rectangular shape with a coercivity of 0.163 T at 10 K, suggesting a ferromagnetic ground state. However, under the influence of strain (1.2% and 1.8% strain), besides the hysteresis loop at a low magnetic field (gray rectangular area in Fig. 5a), a representative step-shaped M-H curve can also be found at a high magnetic field. It indicates a transition from ferromagnetic to antiferromagnetic orders in the part of layers under tensile strain. At the transition field, the magnetic moments in the antiferromagnetic layers are aligned, leading to an increase in magnetization. As shown in Fig. S9, the magnetization (intensity of MCD signal) gradually increases as the magnetic field increase and is fully polarized above

$\mu_0 H = 1\ T$, which is comparable with that in an unstrained state. Under 3% strain, the zero magnetization at a low magnetic field indicates the disappearance of the FM state, leaving antiferromagnetic coupling among adjacent layers. This phenomenon has been observed in $CrI_3$ under hydrostatic pressure, where it originates from the change of interlayer stacking order[11]. For our results, the in-plane tensile strain will induce different stretching lengths on different layers, which also influence the stacking order, to the appearance of antiferromagnetic order in $CrI_3$. Until now, the magnetic properties of $CrI_3$ show the coexistence of ferromagnetic and antiferromagnetic states below 3% tensile strain. Although the appearance of the antiferromagnetic state, we can still evaluate the coercivity of the ferromagnetic state under strain by considering the FM at a low magnetic field (gray area). To better evaluate the coercivity under strain, we conducted a fine measurement at a low magnetic field and summarized the result in Fig. 5b. It shows the coercivity is between 0.16 and 0.17 T in the range from 0 to 1.8% strain. The situation for point 2 is a little different. The initial state is the co-existence of antiferromagnetic and ferromagnetic states due to inevitable strain by dry transfer. As tensile strain increases, although the shape of the M-H curve dramatically changes due to changed interlayer stacking order, the change of coercivity of FM state is limited, from 0.16 T to 0.13 T within 1.8% strain. The small variance may come from the finite perpendicular magnetic anisotropy induced by SOC between I 5p orbitals and their hybridization with Cr 3d electrons[27]. In a word, the change of coercivity in $CrI_3$ is negligible in comparison with that in $VI_3$ under the same tensile strain, which comes from the weak SOC in $CrI_3$. In addition, the $T_c$ for two points evaluated from the temperature-dependent MCD measurement at different strains (Fig. S10 and S11) show a weak dependence on the tensile strain.

**Conclusion**

In conclusion, the coercivity of $VI_3$ flakes under an in-plane tensile strain has been carefully investigated, based on in-situ MCD measurements. We find that coercivity in $VI_3$ flakes can be enhanced from 1.78 T in the unstrained state to 2.60 T under a 2.9 % strain at 10 K, without the influence on $T_c$. In addition, the tunability of coercivity in $CrI_3$ flakes under an in-plane tensile strain is limited. This is mainly related to the strong MAE induced by SOC in $VI_3$ in comparison with the quenching orbital moment in $CrI_3$. The findings of this work open up a novel route for effectively tuning the coercivity based on MAE in 2D magnets.

**Author contributions**

J.D., Y.C., and J.M. conceived the project. X.Z., H.S., X.X., and J.D. designed and performed the experiments. L.W., C.L., and M.H. provided and characterized the samples. J.L. conducted the AFM measurement. Y.C. and J.M. provided theoretical support. All authors discussed the results and co-wrote the paper.

## Acknowledgment

We would like to thank Prof. Haizhou Lu from SUSTech for helpful discussions. J.W.M was supported by the National Key Research and Development Program of China (Grants No. 2021YFA1400400). J.F.D. acknowledges the support from the National Natural Science Foundation of China (11974159) and the Guangdong Natural Science Foundation (2021A1515012316). J.W.M. and L.W. were supported by Guangdong Basic and Applied Basic Research Foundation (No. 2020B1515120100).

## Conflict of Interest

The authors declare no conflict of interest.

## Method：

**Sample growth:**

$VI_3$ single crystals were synthesized by a chemical vapor transport method. The mixture of vanadium powder and iodine pieces with a molar ratio of 1:3.1 was loaded in a silica tube and sealed under a vacuum. The ampoule was then put into a single-zone horizontal tube furnace, with the hot end held at 823 K and the cold end at ambient temperature. After 3 days of vapor transport, millimeter-sized hexagonal dark brown plates were obtained at the cold end, as shown in Fig. S1. The samples are sensitive to humidity. The single-crystal XRD pattern was performed at room temperature on a Rigaku Smartlab-9 kW diffractometer with Cu K$\alpha$ radiation ($\lambda_{K\alpha1}$ = 1.54056 Å). The magnetic susceptibility (M/H) was performed on a Quantum Design (QD) MPMS3 SQUID magnetometer.

**In-plane tensile strain using PDMS:**

We employed thin polydimethylsiloxanes (PDMS) as a stretcher to apply a tensile strain on the layered materials. First, we cut the PDMS into a long strip, and then attached it to a $SiO_2$/Si substrate. After that, we carefully exfoliated $VI_3$ and $CrI_3$ flakes on the PDMS with blue tape, to keep the sample in an unstrained state. To apply a tensile strain on the samples, we first pulled PDMS/substrate apart at one boundary, stretched PDMS with a micro-manipulator, and then put PDMS back to the substrate again to keep the tensile strain. The strain was estimated by measuring the change of length of the samples along the stretching direction. To exactly evaluate the strain, we checked the length change in different positions of samples.

**Magnetic circular dichroism (MCD) system:**

The ferromagnetic behaviors of the $VI_3$ flakes under in-plane tensile strain were measured in an in situ magnetic circular dichroism (MCD) system (Figure S4). We use a 632.8 nm HeNe laser as the light source. After passing a long working distance objective, it is focused into a ∼2 μm spot on the sample mounted on the cold finger within the cryostat. The polarization state of the incident light beam changes between the left-handed and right-handed circular polarization with a frequency of 50 kHz using a photoelastic modulator (PEM). The reflected MCD signal, i.e., the intensity difference between two circularly polarized states, is directed to a silicon detector and recorded by a lock-in amplifier. During measurements, the out-of-plane magnetic field generated from a superconducting loop is scanned within ±3 T. Hence, we can record the MCD signals as a function of the applied magnetic field at different strains. In the MCD mapping, we fix the applied magnetic field at several fixed values and measure the MCD signals as a function of position with ∼1 μm resolution.

**First-principles calculations:**

All calculations were performed by using the Quantum ESPRESSO package[28, 29] based on density functional theory. The electron exchange-correlation effect was described within the generalized gradient approximation (GGA). We used ultra-soft pseudopotentials with a plane-wave energy cutoff of 60 Ry. We utilized a Monkhorst-Pack $k$-mesh of 8×8×1 for monolayer $VI_3$ in the structural optimization but a denser $k$-mesh of 12×12×1 in calculating the electronic and magnetic properties. The structure was fully relaxed until the force on each atom was smaller than 0.001 Ry/Bohr. The vacuum space between adjacent slabs was set to 15 Å. The spin-orbit coupling effect (SOC) was considered for magnetic anisotropy energy (MAE) calculation based on the canonical formulation.[30]

FIGURES AND CAPTIONS

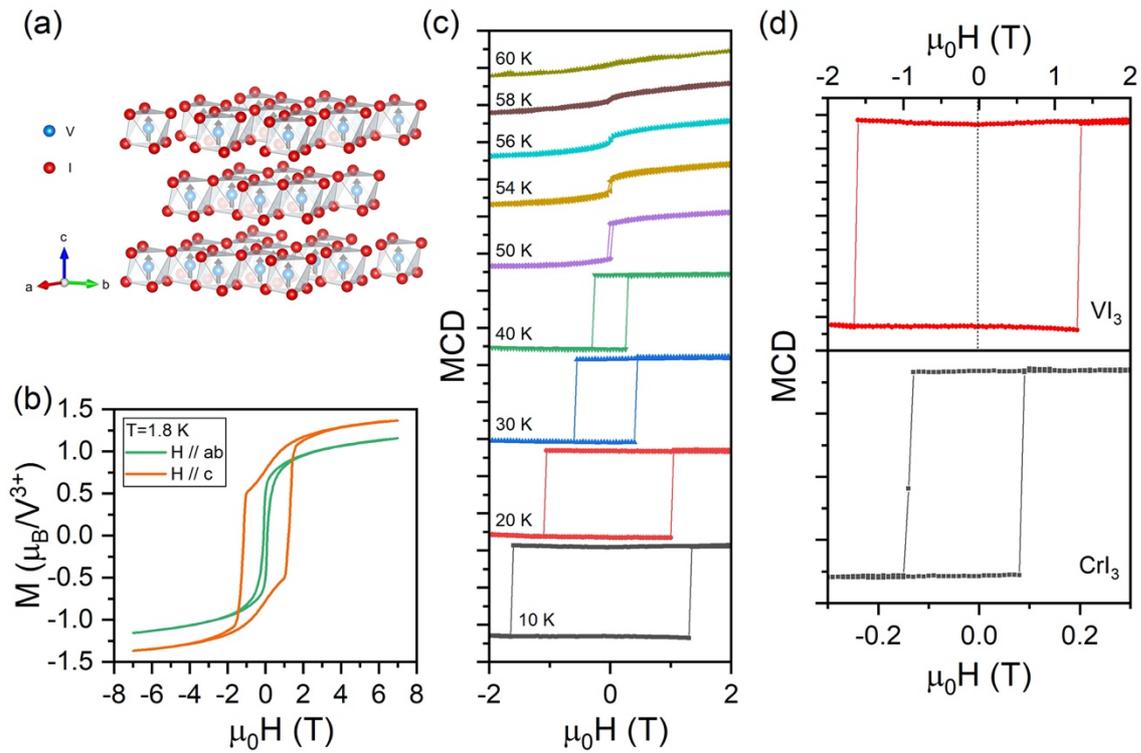

Figure 1. (a) Side view of VI$_3$ single crystal. The arrows indicate that the magnetization is along the c-axis. Blue and red spheres correspond to vanadium (V) and iodine (I) atoms, respectively. (b) Field-dependent magnetization of VI$_3$ single crystal at 1.8 K for H//ab (green curve) and H//c (orange curve), respectively. (c) Temperature dependence of MCD signals in a 27 nm VI$_3$ flake. (d) MCD signals as a function of an applied magnetic field of VI$_3$ and CrI$_3$ flakes at 10 K, respectively.

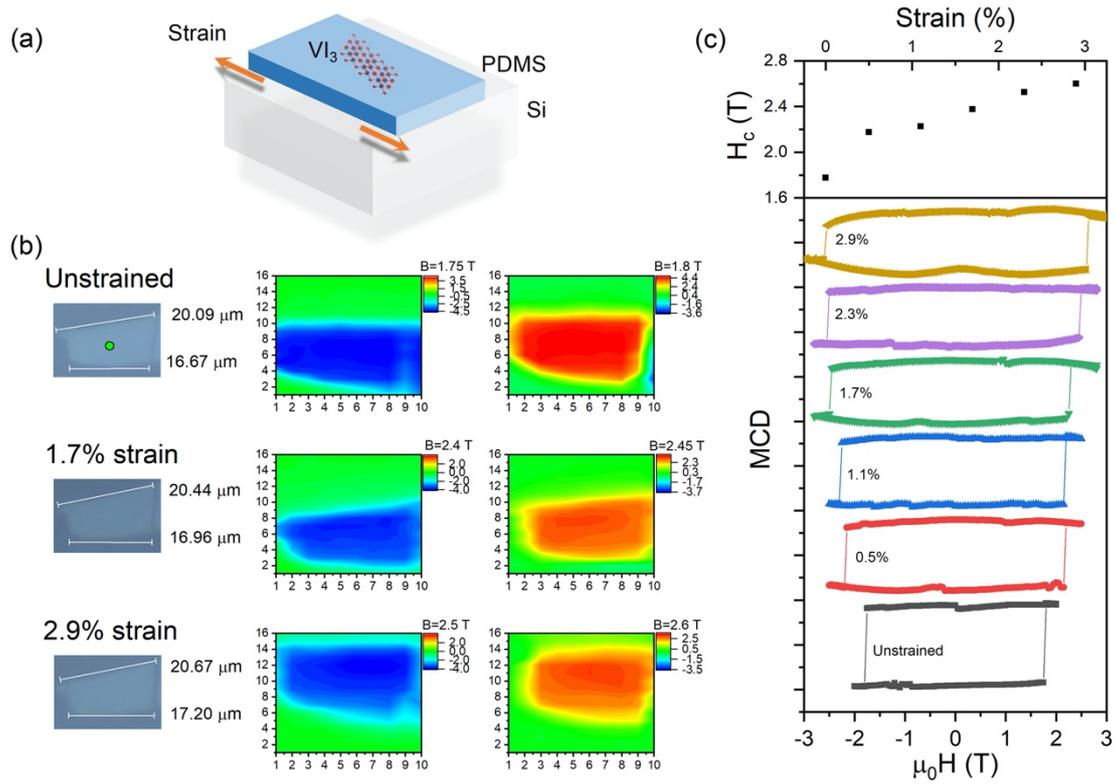

Figure 2. (a) Schematic diagram of applying an in-plane tensile strain in the VI$_3$ flake. (b) Three pictures on left show the unstrained, 1.7%, and 2.9% strain, respectively. Right pictures show the MCD mapping under three representative strains, e.g. 0%, 1.7%, and 2.9% strain. At first, the magnetization is polarized by a negative magnetic field, the mappings are conducted near the positive coercivity. (c) Magnetization as a function of an applied magnetic field in a strain VI$_3$ flake from 0 to 2.9% at 10 K. The upper figure shows the extracted averaged coercivity as a function of applied tensile strain at 10 K.

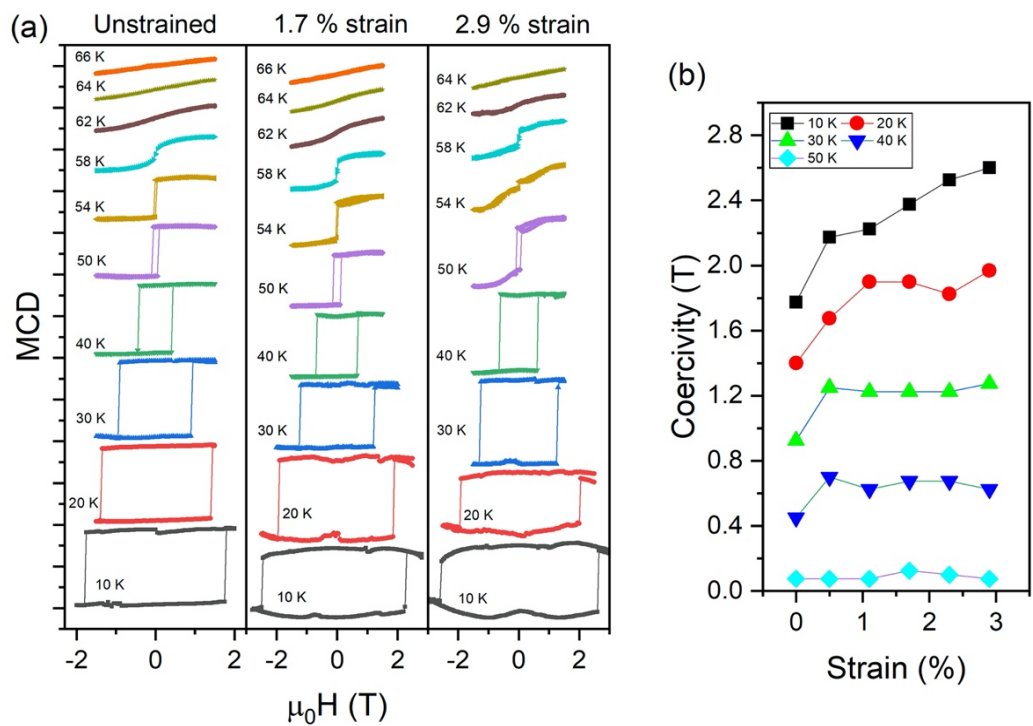

Figure 3. (a) Temperature dependence of MCD signals in the VI$_3$ flake under three fixed strains, namely, unstrained, 1.7%, and 2.9% strain, respectively. (b) Extracted averaged coercivity as a function of strain at five temperatures.

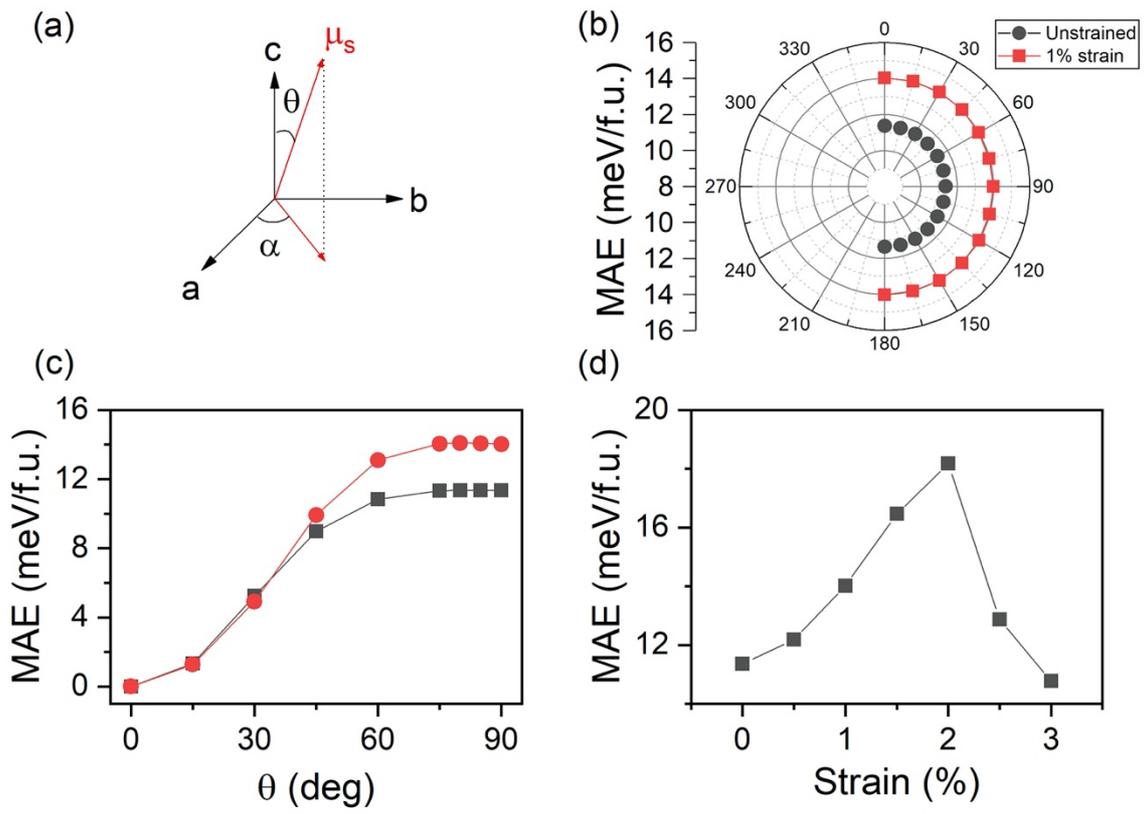

Figure 4. (a) Schematic to show the definition of magnetic moment direction θ and α. (b) The α and (c) θ dependence of MAE in a monolayer $VI_3$. The black and red data stand for unstrained and 1% tensile strained monolayer $VI_3$. (d) The strain-dependent MAE in a monolayer $VI_3$.

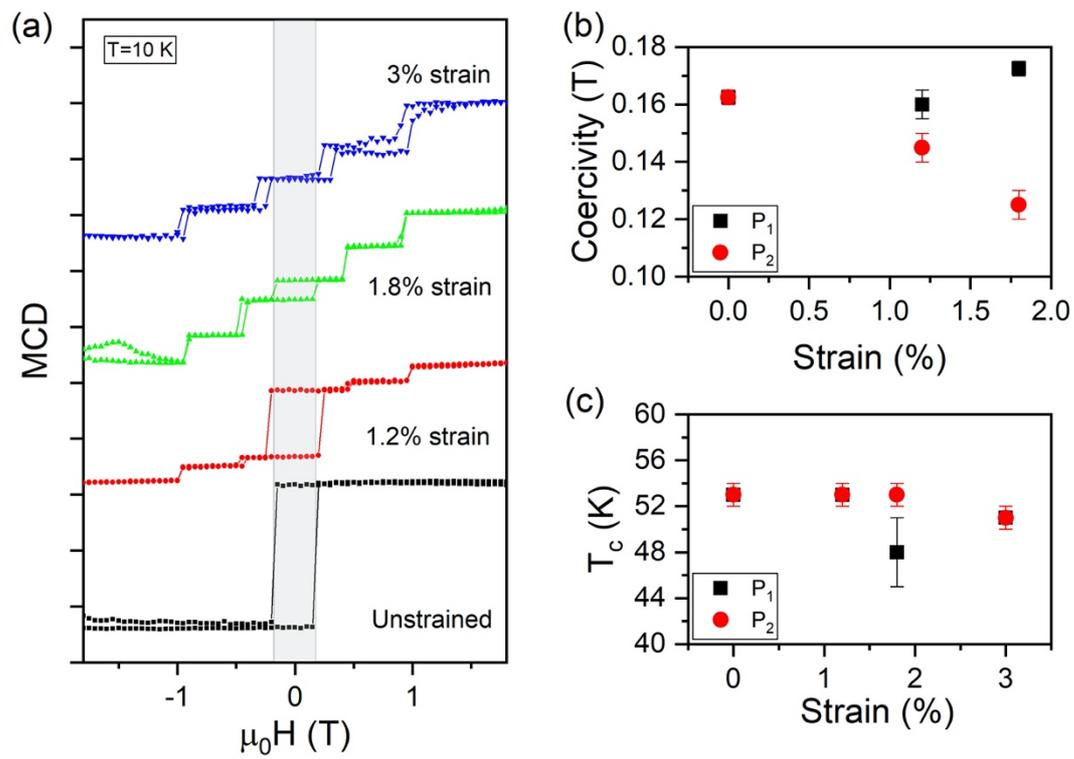

Figure 5. Strain tunability in a thin $CrI_3$ flake. (a) MCD signals as a function of an applied magnetic field at 10 K under several strains, namely 0, 1.2%, 1.8%, and 3%. (b) The extracted coercivity of FM state as a function of tensile strain in two fixed points ($P_1$ and $P_2$) of the $CrI_3$ flake at 10 K. (c) Curie temperature ($T_c$) as a function of tensile strain in $P_1$ and $P_2$.